\documentclass[12pt]{article}
\usepackage{color}
\usepackage{latexsym}
\usepackage{amsmath}
\usepackage{amsfonts}
\usepackage{amssymb}
\usepackage{indentfirst} 
\numberwithin{equation}{section}
\newcommand{\be}{\begin{equation}}
\newcommand{\ee}{\end{equation}}

\newcommand{\mH}{{\mathcal H}}
\newcommand{\mK}{{\mathcal K}}

\newcommand{\mA}{{\mathcal A}}

\title{Niederer's  transformation,  time-dependent oscillators and  polarized gravitational waves
\\
}
\author{
\\ 
\vspace*{0.3cm}
K. Andrzejewski\footnote{Corresponding author, e-mail: k-andrzejewski@uni.lodz.pl} \quad and 
 \quad S. Prencel
\vspace*{0.8cm}
\\
\small  Department of  Computer Science,  Faculty of Physics    and Applied Informatics, \\ \small
 University of  Lodz, Pomorska 149/153,
90-236, Lodz, Poland\\
}
\date{}
\begin{document}
\maketitle 
\begin{abstract}  It is noted  that the Niederer transformation can  be used to find the explicit  relation between  time-dependent linear oscillators, including  the most interesting case  when one of them is  harmonic.  
A geometric interpretation  of  this correspondence  is provided  by     certain subclasses of  pp-waves; in  particular   the ones strictly related to the proper conformal transformations. 
 This observation allows us  to show that the    pulses of plane gravitational  wave exhibiting   the maximal conformal symmetry  are analytically solvable.
 Particularly interesting is the circularly polarized  family for which some aspects (such as the classical cross section, velocity memory effect and  impulsive limit) are  discussed in more detail. 
 The role of   the  additional integrals of motion,  associated with  the conformal generators, is clarified by means of Ermakov-Lewis  invariants. Possible applications to the description of  interaction of electromagnetic beams with matter  are also indicated. 
\end{abstract}
\section {Introduction}
Harmonic oscillators with the  time-dependent frequencies  appear in many physical contexts.  They have been extensively studied  at classical and quantum levels. In particular, the possibility of reducing their dynamics to that of an  ordinary harmonic oscillator by means of  canonical transformations  accompanied with a suitable  redefinition of the time variable was investigated in some detail,   see e.g. \cite{a0a}-\cite{a2}.
\par
The paradigm of such an  approach is  provided by the Niederer transformation  which maps the  free motion onto the half-period motion  of  the harmonic oscillator (also on both levels) \cite{b7a}. The starting point  in the present  paper is the observation  that, once  the Niederer  transformation is formulated at the  Lagrangian level, it can easily  be generalized  to the one relating two  linear oscillators with time-dependent  frequencies  connected by a simple formula (cf. eqs. (\ref{e00c})  and (\ref{e00d}) below). In particular, the latter  can be applied   when the initial oscillator is  harmonic yielding in this way an example of exactly  solvable  time-dependent oscillator. 
\par
In order to gain a deeper understanding of the above described properties of  Niederer's transformation  one can refer to  a geometric  picture. Let us recall that  the Niederer  transformation  has a nice  geometric  interpretation  in terms  of Bargmann spaces  \cite{b7b,a3} and the Eisenhart-Duval lift \cite{a4,a5}.   In the  particular case of a 2-dimensional  oscillator considered  below one obtains the  4-dimensional Bargmann manifold  by adding a new variable $v$ which accounts  for the nontrivial  transformation  properties  of the action  under the Niederer  mapping.  The basic identity  relating  the transformed action  to the initial one  is reformulated as the conformal equivalence  of two metrics on the Bargmann manifolds (see eq. (\ref{e12bb}) below). This  relation  admits the   immediate generalization,   see eq. (\ref{e00e}) below,   which provides a geometric picture  for the discussed case.  
\par
The Bargmann metrics  related by   Niederer's  transformation may be viewed  as describing spacetime in form of (generalized)  plane  gravitational    waves. However, due to the fact that Einstein's equations are not conformally invariant at most one metric corresponds to  a vacuum solution. In spite of this  the Niederer transformation  seems  to be very  useful. It allows, for example,  to simplify  the geodesics equations. To see this  let us first note that, being  conformally  equivalent, both metrics yield the equivalent  equations for null geodesics only.  However,  coming back  to the origin of Niederer's  transformation  we conclude  that the geodesics equations for the transversal variables are of the same form; only the equations corresponding to the longitudinal  variables are not equivalent. On the other hand, due to  the special form of  metrics,  the equation for the  longitudinal  variables  decouple and  they can easily be  solved  once the remaining  ones are solved (cf. eqs. (\ref{e6a})-(\ref{e7})).
\par
{In view of the above the Niederer transformation can be useful in analysis of some  properties of plane gravitational waves. This is interesting for several reasons. First, the  plane gravitational waves are solutions to the nonlinear  Einstein equations. Second, due to the vast  distance from the source they can  approximate an arbitrary gravitational    wave in a neighbourhood of the detector. Moreover,  the Penrose limit of spacetimes yields the (generalized)  plane gravitational waves. In consequence,   such  waves  can be used to model and   study  some more complex  problems appearing in physics of the  gravitational waves.   
 Among others the gravitational memory is becoming an increasingly important topic,  not only because of its potential observability in gravitational waves from
astronomical sources \cite{b3d,b3e}, but also for its importance in theoretical issues in quantum gravity \cite{b4b}. The memory  effect can be simply stated as 
displacement (or a residual velocity) noted by freely falling detectors, due to the passage of a pulse of gravitational
radiation.   This may take the form of a fixed change in position (permanent displacement)  or a constant separation velocity. 
A permanent change in the Minkowski spacetimes which  exists before the arrival of the pulse and
after its departure  was noted in Ref.  \cite{x1} and  considered  in the context of linearized   gravity  in Refs. \cite{x2,x3}. Next it  was   
extended to the full nonlinear theory of general relativity \cite{x4}; moreover,  it turns out that there is a ``nonlinear" contribution to the memory effect,  due to the effective stress energy of the gravitational waves  transported to null infinity (the latter  effect can be  much larger than
the formerly known ``linear" one). Following this idea, it was shown  \cite{x5,x6,x7} that there is a contribution to the memory from other
particles having zero rest mass and hence distinguished between the linear and non-linear contributions as ordinary
and null memory; the stress energy travels to null infinity in the latter case only. 
Apart from four-dimensional asymptotically flat spacetimes, the memory effect  has been studied in other  spacetimes as well as in higher
dimensions \cite{x8a}-\cite{x8d}; there are also   analogous  effects for electromagnetic fields \cite{x9}. Various extensions of this effect have also been studied in modified gravity and massive gravity theories
\cite{x9b}. For the case of the plane gravitational waves such considerations trace back to the  work \cite{b6c}. However, recently they   gained a new impact due to the  direct observations of the gravitational waves and  its  role in    understanding   inequivalent ground states (vacua)  important for certain aspects of  soft gravity  (asymptotically flat spacetimes before the arrival of
the pulse and after its departure are inequivalent and related via transformation which   do not tend to the identity at infinity), see e.g. \cite{b2c}-\cite{x0c}. Especially interesting is also  the observation   that memory is encoded in the Penrose limit of the original gravitational wave spacetime recently noticed in \cite{x0c}; for null congruences, the Penrose limit is a plane
wave. Such  analysis enhances the range of applications of existing studies involving geodesic deviation and memory in plane wave spacetimes.}
\par 
{
In the current study  the   reasoning based on  the Niederer transformation  is applied  to some special    vacuum metrics  distinguished by the existence of    proper (special) conformal generators \cite{b6a,b6b}.   They describe some linearly and circularly polarized  plane gravitational waves.  Such waves seem to be of some interest   because 
circularly polarized gravitational  waves  may  be produced  by coalescing black holes,  neutron star merger or observed from the  astrometric data   \cite{b1h,b1i,b4c}; (the linearly polarized  case   can be also  interesting due  to its relative simplicity).
 Using the above mentioned approach  we directly obtain  the solutions of    the geodesics equations for the  plane gravitational waves  under  consideration  and explicitly discuss  some phenomena 
such as  velocity memory effect,  classical  scattering  and focusing.  Furthermore, by taking an appropriate  limit  we notice that    they can be used to model   impulsive gravitational waves    \cite{b5e}-\cite{b5j} with the  Dirac delta profile. 
{
Moreover, the role and meaning  of  the   integrals of motion  associated with  the conformal generators is clarified by means of Ermakov-Lewis  invariants.
}
}
\par
The paper is organized as follows. In Sec. 2 we describe   the Niederer transformation in the above motioned general approach and apply it to   obtain  solutions of    an isotropic   time-dependent oscillator related to  a special, conformally flat, null fluid solution of Einstein's equations.  Next, in Sec. 3  we explicitly discuss the interaction of the  particle with the plane gravitational wave pulses defined by the maximal, 7-dimensional, conformal symmetry; moreover, we give  possible applications  of the  Niederer  transformation  to the pp-wave spacetimes. The meaning  of the integrals of motion associated with conformal generators is discussed in Sec. 4. Finally,  Sec. 5 is devoted   to some  conclusions. In particular, we briefly  mention the possible applications  of  the results obtained here  in a construction  of electromagnetic backgrounds  important  for the light-matter interaction. This is  motivated by  recent results on the classical double copy approach \cite{b8a}-\cite{b8d}  which enable one  to build  some  electromagnetic fields from  solutions to the Einstein equations.  Such an  approach, when applied to the families of plane gravitational  waves discussed here,   may lead to   electromagnetic beams similar to the ones appearing in the context of singular optics \cite{b9,b9a} and  intense lasers \cite{b10a,b10b}. 
\section{Niederer's transformation}
{In this section we recall   the   notion  of the Niederer transformation and  subsequently  we   extend it   to the   case  of   time-dependent linear  oscillators; we also  notice that  a   geometric picture of such  a situation  is a  very convenient framework  to  analyse the  geodesic (deviation)  equations.}
It  is well known that  the  free motion can be  mapped into the harmonic  one (for our purpose we consider  2-dimensional case).  Namely, the so-called  Niederer transformation\footnote{For   further convenience  we adopt, non-standard,   $u$-notation  while  bold indices refer to 2-dimensional vectors. Let us also   note  that there exists  a  hyperbolic  counterpart of  Niederer's transformation leading to the repulsive case.}
\be
\label{e12}
\begin{split}
u&=\epsilon\tan(\tilde u), \\
{\bf x}&=\frac{\epsilon \tilde {\bf x}}{\cos(\tilde u)},\\
\end{split}
\ee
relates the free motion $\overset{..}{{\bf x}}=0$ on the whole real axis ($-\infty<u<\infty$) to the half of period motion ($-\frac\pi 2<\tilde u<\frac \pi 2$) of the  attractive harmonic motion $\tilde{\bf x}''=-\tilde {\bf x}$; here  dot and prime  refer to the derivatives with respect to $u$ and $\tilde u$, respectively. On the level of action functional this basic property of the Niederer transformation    is expressed by the following identity 
 \be
 \label{e0}
\frac {1} {2}  \dot{\bf  x}^2 du= \frac 1 2(  \tilde{\bf  x}'^2 -\tilde{\bf x}^2)\epsilon d\tilde u+\left(\frac{1}{2} \tan(\tilde u)\tilde {\bf  x}^2\right)'\epsilon d \tilde u,
\ee 
Furthermore, this equivalence continues to   hold at the quantum level  \cite{b7a}; namely, if $\psi({\bf x}, u)$ is a solution to the Schr\"odinger equation for the free particle then 
\be
\label{e00b}
\phi(\tilde {\bf x},\tilde u)= \frac{e^{-\frac{i\epsilon}{2}\tan(\tilde u)\tilde {\bf x}^2}}{\cos(\tilde u)}\psi\left (\frac{\epsilon\tilde {\bf x}}{\cos(\tilde u)},\epsilon\tan(\tilde u)\right),
\ee
is a solution to the Schr\"odinger equation for the harmonic oscillator.  The structure of the above map 	    is as follows:  first, the arguments of the wave function are replaced by the appropriate functions of the new ones according to the classical formulae; then the two factors are added: the first one accounts for  the proper normalization while the other, the phase one,  is exactly equal to the function which enters the total time derivative in the transformation rule  (\ref{e0}). 
 \par 
 Of course,  the above  observations  have a local character  and their domain of validity must be carefully analysed.  However,  they reflect a similarity between  both systems  and bring immediately some useful information. 
 Various local quantities (like symmetry generators)  can be directly related; this  concerns  even the global  ones (for instance, Feynman propagators) if sufficient care is exercised (see, e.g.  \cite{b7a,b7c,b7d}).   In particular,   the  maximal symmetry groups of both systems   are isomorphic. Moreover, one obtains  the explicit relation between their solutions  which  enables for a  more detailed analysis  \cite{b7e}. 
\par 
Now, let us note that adding the  term $\frac{1}{2}{\bf x}\cdot H(u){\bf x}du$ (where  $H$ is, without loss of generality,  a symmetric matrix)  to the  both sides  of  identity (\ref{e0}) one obtains  the following  relation 
\be
\label{e00c}
\frac {1} {2}  (\dot{\bf  x}^2 +{{\bf x}\cdot H(u)}{\bf x})du= \frac 1 2(  \tilde{\bf  x}'^2 +\tilde {\bf x}\cdot \tilde H(\tilde u)\tilde{\bf x})\epsilon d\tilde u+\left(\frac{1}{2} \tan(\tilde u)\tilde {\bf  x}^2\right)'\epsilon d \tilde u,
\ee 
where 
\be
\label{e00d}
\tilde H(\tilde u)=\frac{\epsilon^2H(\epsilon \tan(\tilde u))}{\cos^4(\tilde u)}-I,
\ee
which implies  that  the Niederer  mapping  connects two linear  oscillators with  time-dependent frequencies. 
In particular, if   we introduce a matrix $G$ such that 
\be
\label{e00k}
 H(u)=\frac{a}{(\epsilon^2+u^2)^{2}}G(u),
\ee
then,  by virtue  of eqs. (\ref{e00c}) and (\ref{e00d}),  the equations of motion 
\be
\label{e00f}
\overset{..}{{\bf x}}=H(u) {\bf x},
\ee
transform into the following ones
\be
\label{e00g}
\tilde {\bf x}''=\tilde H(\tilde u)\tilde {\bf x}= \left(\frac{a}{\epsilon^2} G(\epsilon \tan(\tilde u))-I\right)\tilde {\bf x}.
\ee
{In consequence  we transformed  a linear oscillator (\ref{e00f}) into another one  described by eq. (\ref{e00g}); of course the advantage of such result is when the latter one  is more tractable  or interesting. In what follows we give some example of such situation.}
\par Let  us consider  a   special   choice    of  the matrix $G$ (equivalently $H$) namely,   $G=I$;  then (\ref{e00f})  and (\ref{e00g})    imply that    the (non-singular) time-dependent  linear  oscillator
\be
\label{e0a}
\overset{..}{\bf x}=\frac{a}{(u^2+\epsilon^2)^2}{\bf x},
\ee
where $a,\epsilon$  are non-zero real numbers\footnote{For  $a=0$ this reduces to the standard realization.  For  the singular cases: $\epsilon=0$ the transformation is of the form     $u=\tilde u^{-1}$, ${\bf x}=\tilde{\bf x}\tilde u^{-1}$;  if the denominator contains $u^2-\epsilon^2$ then one can use the hyperbolic counterpart of Niederer's map.}, is mapped under (\ref{e12})    to
\be
\label{e0b}
\tilde{\bf x}''=\left(\frac{a}{\epsilon^2}-1\right)\tilde{\bf x},
\ee
i.e. to a part of  the motion of   the harmonic oscillator or even,   when $a=\epsilon$,  to  the free motion (on the   interval $(-\pi/2,\pi/2)$ only,    in contrast to standard case), in the agreement with the general theory,  see e.g. \cite{a1,a2,b7b,a5}. In consequence, the Niederer transformation  immediately yields  the explicit form   of  the solution to eq.  (\ref{e0a}) (cf. eq.  ({\ref{e10})  below for $i=1$). 
Moreover, since eq. (\ref{e0b})  describes the isotropic harmonic oscillator one can further  use the inverse of  the Niederer transformation (with appropriate parameters)  to pass  at least locally   to the  free motion. 
\par Of course, a  similar reduction (in general to the anisotropic case)  holds for an arbitrary  constant matrix $G$  (especially interesting is  the   case when   $\textrm{tr(G)}=\textrm{ tr}(H)=0$).  This observation    not only  enables one    to  understand better  some properties of the   time-dependent  oscillators   but   it     has also  a  natural geometric counterpart. 
 \par 
 { To this end let us recall that the Niederer  transformation received   a  nice geometric interpretation in terms of  the  Bargmann manifolds \cite{b7b,a3} (see also \cite{a4,a5} and references therein).  Namely,} it is viewed as  a chronoprojective transformation (a kind of conformal map;   in the context of plane gravitational waves see  the  quite recent Reference    \cite{xp})  between the  Bargmann spacetimes  corresponding to the free $(u,{\bf x},v)$ and harmonic  $(\tilde u,\tilde {\bf x},\tilde v)$ motions \cite{b7b}.  In this approach  the term  which appears in form of  total time derivative in  (\ref{e0}) (or the phase factor in (\ref{e00b}))   enters the transformation rule  for  the additional, $v$ and $\tilde v$, variables parametrizing  the Bargmann manifolds, i.e. 
 \be
 \label{e12b}
 v=\epsilon\tilde v-\frac {\epsilon \tan(\tilde u) }{ 2}\tilde {\bf x}^2.
 \ee
Eq. (\ref{e12b})  together with (\ref{e12})  lead to the following identity
 \be
 \label{e12bb}
 {\bf  dx}^2 +2dudv=\frac{\epsilon^2}{\cos^2 (\tilde u)}(d\tilde{\bf  x}^2+2d\tilde u d\tilde v -\tilde {\bf x}^2d\tilde u^2),
 \ee
 between   the  free  and half-oscillatory period  Bargmann spaces (see also \cite{a6,a6b}). 
 Furthermore, the  counterpart of the identity   (\ref{e00c})   reads 
\be
\label{e00e}
g\equiv {\bf x}\cdot H(u){\bf x}du^2+2dudv+d{\bf x}^2=\frac{\epsilon^2}{\cos^2(\tilde u)}(\tilde {\bf x}\cdot \tilde H(\tilde u)\tilde {\bf x}d\tilde u^2+2d\tilde ud\tilde v+d\tilde {\bf x}^2)\equiv \frac{\epsilon^2}{\cos^2(\tilde u)}\tilde g,
\ee
where $H$ and $\tilde H$  are connected by eq. (\ref{e00d}).  The metric $g$  (and, consequently, $\tilde g$)   described in  (\ref{e00e}) belongs to   a subclass of   pp-waves, the so-called   generalized plane gravitational  waves. In general, they do not satisfy the vacuum Einstein equations; only scalar curvature is zero while  the source is a null fluid (i.e. they are   solutions to Einstein's field equation with the energy-momentum tensor  describing some kind of   radiation).  The weak energy condition implies  $\textrm {tr}(H)\leq 0$. For $\textrm {tr}(H)= 0$ the metric   $g$   satisfies  the vacuum Einstein equations  and  describe  the  plane gravitational   wave (exact gravitational wave).  
\par 
{Now,  let us concentrate on the geodesic equations. To this end let us recall that the  geodesic equations for the metric $g$}, appearing in  eq. (\ref{e00e}),  are  of the form 
\begin{align}
\label{e6a}
\overset{..}{{\bf x}}&=H {\bf x},\\
\label{e6b} \overset {..}{v}&=-\frac12 {\bf x}\cdot  \dot{H}{\bf x}-2 {\bf x}\cdot H	\dot {\bf  x}.
\end{align}
Moreover   eq.  (\ref{e6a}) coincides with  the deviation equation.  Eq. (\ref{e6b}) can be integrated  to the form 
\be
\label{e7}
v(u)= -\frac  12 {\bf x}\cdot \dot {\bf x}+C_1u+C_2,
\ee
where ${\bf x}$ is a  solution of (\ref{e6a}).   
\par
{Let us now assume that  we  are interested  in solving  the geodesic  equations for the metric  $g$};  then by applying  the Niederer transformation  one can associate with $g$ the  new  metric  $\tilde g$  conformally related  to $g$  (see eq. (\ref{e00e})). Due to the fact that  Einstein's equations are  not conformally invariant  at most only  one of the  metrics $g$ and $\tilde g$  describes  the vacuum  solution. However, we are interested  in geodesics. Even though the  geodesics  equations  for $g$ and $\tilde g$ are not  equivalent  (except  those for null geodesics), it follows  from the reasoning  underlying  the  derivation of Niederer's transformation that  the equations  for ${\bf x}$ and $\tilde{\bf x}$ are equivalent.  The equations  for ${\bf x}$  transform into the following ones 
\be
\label{e7a}
\tilde{\bf  x}''=\tilde H (\tilde u)\tilde {\bf x},
\ee
which may be simpler in some cases.  The non-equivalence of geodesic equations for  $g$ and $\tilde g$  is reduced  to  the non-equivalence  of equations  determining  $v$ and $\tilde v$. However,  the latter can be easily  solved 
(cf. eq. (\ref{e7}))  once $\bf x(u) $ (or $\tilde {\bf x} (\tilde u)$) are known.   In this  way we are left with  eqs. (\ref{e7a})  which can be more tractable than  those for the initial metric. 
\par 
{Let us  apply the above reasoning;  first, for  the metric $g$  with the profile   $H(u)$ given by (\ref{e00k})  and   $G=I$}. Then  one obtains the conformally flat metric $g$; it is  worth to notice that this metric is distinguished  in the conformally flat subclass of generalized plane   waves,  it is the only  one  which admits the   so-called  proper special conformal vector field (see  \cite{b6a} for more  details). Furthermore, the  transverse geodesics equations for  $g$ and $\tilde g$    are given by the formulae (\ref{e0a}) and (\ref{e0b}), respectively; thus they   can be explicitly solved (and consequently (\ref{e6b})). 
Let us also note  that  if  $H(u)$ is proportional to the identity then the  same concerns the matrix $\tilde H(\tilde u)$. Therefore  we do not leave out the conformally flat subclass of the null fluid solutions (in   the  Eisenhart-Duval lift language  the class of  isotropic oscillators, including also the free motion). 
\par 
 If $g$ describes  the plane gravitational  wave  the situation is   quite different;   then $\tilde g$  is  not  an exact    gravitational wave but  a, non-vacuum and non-conformally flat, null fluid solution (in     the Eisenhart-Duval lift language,   anisotropic oscillators are transformed into  anisotropic ones).   However,  again for some special  plane  gravitational waves  the geodesic equations for $\tilde g$ (and consequently for $g$ too) can be explicitly solved (see below).  
 \section{Polarized plane  gravitational waves}
\subsection{The general discussion} 
\label{sub1}
{As we noted in the previous section  the Niederer map  can be used to rewrite  the  geodesic equations in, perhaps,  a simpler form.  However, a priori it is not clear  for which profiles of plane gravitational waves such a trick  should work; the geodesic equations do not distinguish any profiles.  Usually some hints concerning the integrability provides  the isometry group; however,  it is well known that the generic dimension of the  isometry group of the gravitational plane waves   is  five; there are  two exceptional classes  with the six-dimensional isometry group, they are extensively studied in the literature:   a kind of periodic waves (applied, for example,  to  describe  the  gravitational wave  which   is  a sandwich  between two Minkowskian regions) and   not  geodesically complete  class (used   in the context of  the Penrose limit \cite{b6f}).   Thus, in what follows  we  consider  the conformal symmetry}. 
\par 
To this end let us recall   that the conformal group of a non-conformally flat spacetime is at most 7-dimensional and the maximal dimension is  rather rarely attained. If we restrict  ourselves to  the vacuum solutions then there are only three classes of  metrics exhibiting  the 7-dimensional  conformal symmetry  \cite{b6a,b6b}  (see also \cite{b6c,b6d,b6e});  all of them describe  plane  gravitational waves. Two of them coincide with the two above mentioned classes.    We shall discuss  the third one which   consist of the linearly and circularly polarized families and,   in contrast to the previous  classes,  exhibits a proper  (non-homothetic)  conformal vector field. Since we are   interested  in  continuous  gravitational pulses we concentrate on    the geodesically  complete  cases (with the non-singular metrics). Then the  first, linearly polarized, family  is defined (up to $u$ translation)   by the     metric $g^{(1)}$ with     the profile 
\be
\label{e8}
H^{(1)}(u)=\frac{a}{(u^2+\epsilon^2)^2} G^{(1)}(u), \qquad  G^{(1)}(u)=
\left(
\begin{array}{cc}
1&0\\
0&- 1
\end{array}
\right),
\ee
where $ \epsilon>0$ and $a$ is an arbitrary number (excluding the trivial Minkowski case and redefining, if necessary, $x^1$ and $  x^2$  one can achieve   $a>0$). Moreover, let us note that  taking $a\sim \epsilon^3$  one obtains the model of  an impulsive gravitational wave with  the Dirac delta profile (as $\epsilon$ tends to zero).  
\par 
The second family $g^{(2)}$  provides an example of   the circularly polarized plane  gravitational waves. It is defined by the following profiles 
\be 
\label{e9}
H^{(2)}(u)=\frac{a}{(u^2+\epsilon^2)^2}G^{(2)}(u), \qquad G^{(2)}(u)=
\left(
\begin{array}{cc}
\cos(\phi(u)) &\sin(\phi(u))\\
\sin(\phi(u)) & -\cos(\phi(u))
\end{array}
\right),
\ee
where 
\be
\label{e9a}
\phi(u)=\frac{2\gamma}{\epsilon}\tan^{-1}(u/ \epsilon),
\ee
and $\epsilon, \gamma>0$  and  $a$ can be chosen as above  (for $\gamma=0$ eq. (\ref{e9})  reduces to the previous case; however, for both physical and mathematical reasons we will consider the linear and circular polarizations  separately).
\par 
Now, by virtue of (\ref{e00d}) and (\ref{e00e})     (see   also  \cite{b6a}),   the   Niederer transformation  given by eqs. (\ref{e12}) and (\ref{e12b})     leads to the following relation 
\be
\label{e14a}
g^{(1,2)}=\frac{\epsilon^2}{\cos^2(\tilde u)}\tilde g^{(1,2)},
\ee
where  the metrics $\tilde g^{(1)}$ and $\tilde g^{(2)}$ are  defined by the profiles 
\be
\label{e14}
\tilde H^{(1)}(\tilde u)=
\left(
\begin{array}{cc}
\frac{a}{\epsilon^2}-1 &0 \\
 0& -\frac{a}{\epsilon^2}-1
\end{array}
\right), \quad 
\tilde H^{(2)}(\tilde u) =
\left(
\begin{array}{cc}
\frac{a\cos(\frac{2\gamma}{\epsilon}\tilde u)}{\epsilon^2}-1 &\frac{a\sin(\frac{2\gamma}{\epsilon}\tilde u)}{\epsilon^2}\\
\frac{a\sin(\frac{2\gamma}{\epsilon}\tilde u)}{\epsilon^2} & -\frac{a\cos(\frac{2\gamma}{\epsilon}\tilde u)}{\epsilon^2}-1
\end{array}
\right),
\ee
respectively. Let us stress that, in contrast to $H^{(1,2)}$,   $ \textrm{tr}(\tilde H^{(1)})=\textrm{tr}(\tilde H^{(2)})=-2<0$;  thus  $\tilde g^{(1)}$ and $ \tilde g^{(2)}$ describe, non-vacuum,   null fluid solutions.  
\par 
{Before we go further let us recall that,   in the previous section, we noted that    the conformally flat metric with  $G=I$ is   related to an $u$-dependent  isotropic oscillator. In view of the above,  the first equation in  (\ref{e14a})} provides the  geometric interpretation  of  the Niederer transformation  relating  the   time-dependent  oscillator (described by $H^{(1)}(u)$)  to the   anisotropic harmonic  one (described by $\tilde H^{(1)}(\tilde u)$). 
\par 
{ Now, on the basis of  the above  observations  we  to    give  a   detailed discussion  of some phenomena for  the considered   families of polarized  gravitational pulses.}The starting point is the observation,   due to (\ref{e00f}) and (\ref{e00g})    the  transversal  parts of geodesic equations    for the metrics $g^{(1)}$ and $g^{(2)}$ take  the following form 
\be
\label{e14b}
\tilde{\bf  x}''=\tilde H^{(1,2)}(\tilde u)\tilde {\bf x}.
\ee
Thus in order to discuss the behaviour  of a test particle under the plane  gravitational  waves $g^{(1)}$ and $g^{(2)}$ it is sufficient to consider  the above equations with the null fluid  spacetimes  $\tilde g^{(1)}$  and $\tilde g^{(2)}$,   and  then go back to the original variables ${\bf x} $ (as well as to the  $v$ coordinate, see  (\ref{e7})); {this is our main idea}.
\par 
For  $g^{(1)}$,  the  above procedure gives    the  solutions of  geodesics equations discussed   in \cite{b7}; namely,  in the  transverse direction, for  $a<\epsilon^2$,   we have  
\be
\label{e10}
 x^i(u)=C_1^i \sqrt{u^2+\epsilon^2}\sin(\sqrt{\Lambda_i}\tan^{-1}( {u}/{\epsilon})+C_2^i),
\ee
where 
\be
\label{e11}
\Lambda _i={1+(-1)^i\frac{a}{\epsilon^2}} \qquad  i=1,2
\ee
(for $a>\epsilon^2$  the solution $x^2(u)$ does not change  but    $x^1(u)$ is given  by  replacing the   trigonometric functions their  hyperbolic counterparts).
This  allows us   to  explicitly  analyse  the focusing and singularity problems. More precisely, imposing the initial conditions 
\be
\label{e21b}
\dot{\bf  x}(-\infty)=0,\qquad 
\ee
supplied by  the following  one
\be
\label{e21c}
{\bf x}(-\infty) ={\bf x}_{in},
\ee
one concludes  that  only  the  second component $x^2(u)$ exhibits focusing  at one point 
\be
u_0=-\epsilon\cot(\frac{\pi}{\sqrt{\Lambda_2}})>0.
\ee
However, for parameters $a>\epsilon^2$ the situation  changes drastically; namely,    for sufficiently large $\epsilon$   there can be several focusing points  \cite{b7}.   
This fact    complicates   the singularity analysis. More precisely, so far we  discussed the   plane gravitational  waves in  the  so-called Brinkmann coordinates,  where   both   the wave and    geodesics are global (the metric is non-singular).  The Brinkmann coordinates   cover the whole  plane wave spacetime by a single chart. However, the  plane gravitational  waves   are frequently discussed  in the so-called  Baldwin-Jeffery-Rosen   (BJR) coordinates (see, e.g.,  \cite{b12a,b12b} and references therein).  The BJR coordinates, in contrast to the  Brinkmann ones,  are  typically
not global, exhibiting  $u$ coordinate singularities.  This fact is reflected  in  the transformation  rule between  the  both    coordinates. Namely, only a piece of the  Brinkmann  manifold  can be covered by  the BJR coordinates  and consequently at least  two BJR  maps are needed to  completely  describe  the interaction (scattering) of  particle with the  plane gravitational  waves. 
The definition of the  BJR coordinates    is  related to the transverse part of geodesic equations  (see eq. (\ref{e19a}) and below)  and the  minimal number of charts to cover the whole Brinkman manifold is strictly related to the number of  focusing points.   Moreover, a deeper insight into the structure  of BJR coordinates can  be profitable since the  BJR coordinates  seem to be of some importance  in    understanding   inequivalent ground states (vacua) and, consequently,  certain aspects of  soft gravity \cite{b2d}.
{\subsection{The circularly polarized case }} 
In view of the above  discussion it would be interesting to  analyse analytically the behaviour of  geodesics  in  the,  physically more  interesting,  case of      the  circularly polarized gravitational waves.  In what follows we show that  this is actually   possible for  the plane gravitational wave  $g^{(2)}$ (equivalently $\tilde g^{(2)}$,  by  virtue of  Niederer's transformation).  To this end   we introduce the new coordinates ${\bf y}$ (see, e.g.,  \cite{b13b,b14} and references therein) 
\be
\label{e2}
{\bf\tilde  x}=R(\tilde u){\bf y},
\ee
where 
\be 
\label{e2b}
R(\tilde u)=
\left(
\begin{array}{cc}
\cos(\omega \tilde u)&-\sin(\omega \tilde u),\\
\sin(\omega \tilde u)&\cos(\omega \tilde u),\end{array}
\right) , \qquad \omega=\frac{\gamma}{\epsilon}.
\ee
Then,  the metric $\tilde g^{(2)}$   in terms of $y$'s takes the form 
 \begin{align}
 \tilde g^{(2)}&=\left(\frac{a}{\epsilon^2}+\omega^2-1)(y^1)^2+(-\frac{a}{\epsilon^2}+\omega^2-1)(y^2)^2\right)(d\tilde u)^2\\
&+2\omega(y^1dy^2-y^2dy^1)d\tilde u+(dy^1)^2+(dy^2)^2+2d\tilde ud\tilde v,
 \end{align}
 which implies the following geodesic equations
 \be
 \label{e17}
 \begin{split}
(y^{2}) ''  +2\omega (y^1)'+\Omega_- y^2&=0,\\
(y^{1})''   -2\omega (y^2)' +\Omega_+ y^1&=0,
 \end{split}
 \ee
 where 
 \be
 \label{e18}
 \Omega_\pm=1-\omega^2\mp \Omega, \qquad \Omega=\frac{a}{\epsilon^2}.
 \ee
 The above set of equations can be explicitly solved, although the general form of the  solution is rather complicated  \cite{b15a,b15b}. However, let us note that  any solution is a  combination of   trigonometric  functions only or both hyperbolic (or linear)  and trigonometric  ones   depending on the  values of parameters. Indeed, the roots of the characteristic   determinant (for  the functions $y^{1,2}=A^{1,2}e^{it\tilde u}$) read
\be
\label{e19}
\begin{split}
t_{1,2}=\pm\sqrt{1+\omega^2+\sqrt{4\omega^2+\Omega^2}},\\
t_{3,4}=\pm\sqrt{1+\omega^2-\sqrt{4\omega^2+\Omega^2}},
\end{split}
\ee
 thus some  ${\bf \tilde x}$ trajectories remain  periodic and bounded, while others can  spiral outward, cf.  \cite{b14}.
{Thus,  by virtue of   eq.  (\ref{e7}), we can   conclude  that  the circularly polarized case (\ref{e9})  is  also explicitly solvable. This fact  allows us to get insight into some phenomena, such as    the  memory effect,  focusing  and   singularity problems of the BJR coordinates  as well as to  compute the classical cross sections.}
\par
To this end let us recall that the heart    of these  considerations is the matrix $P(u)$  satisfying the  matrix Sturm-Liouville  differential equation
\be
\label{e19a}
\overset{\cdot\cdot}{P}=HP,
\ee
with the  constraint 
\be
\label{e19b}
\dot{P}^TP-P^T\dot P\equiv 0.
\ee
Since the condition (\ref{e19b})  is stable  against evolution it is sufficient  to impose it at   one point.
Now let us assume  that ${\bf x}_1(u)$ and ${\bf x}_2(u)$ are the  solutions to eqs.  (\ref{e6a}) with the  initial   conditions (\ref{e21b})  and the following ones\footnote{We assume that the gravitational wave vanishes  sufficiently fast at  null infinities, see also  \cite{b7}.}
\be
\label{e19c}
{\bf x}_{1,2}(-\infty)={\bf e}_{1,2},
\ee
where  ${\bf e}_{1}$ and ${\bf e}_{2}$ form  the   2-dimensional  canonical base.
Then the matrix 
\be
\label{e19d}
P_{in}(u)=\left({\bf x}_1(u),{\bf x}_2(u) \right),
\ee
related to the initial conditions (\ref{e19c}), is a solution to eq.  (\ref{e19a}) and, by  (\ref{e19c}),  satisfies  the constraint (\ref{e19b}).  As a result, the solutions $\bf{ x}(u)$  to eqs. (\ref{e6a}) with  the the initial conditions (\ref{e21b}) and (\ref{e21c})  can be written as follows 
\be
\label{e19r}
{\bf x}(u) =P_{in}(u){\bf x}_{in}.
\ee 
Thus $\dot {\bf x}(u)=\dot P_{in}(u){\bf x}_{in}$  and consequently the Jacobian $J$ of the  transformation  from the final velocities $\dot{\bf x}(\infty)$  to initial positions ${\bf x}_{in} $  reads
\be
\label{e19k}
J=\frac{1}{\det (\dot P_{in}(\infty))}.
\ee
This Jacobian  is strictly related to the classical cross section (see  \cite{b12b}), 
\be
\label{e19i}
d \sigma_{classical}=dx^1_{in}dx^{2}_{in}=|J|d\dot x^1({\infty})d\dot x^2({\infty})=\frac{|J|}{p_v^2}d\dot p^1({\infty})d\dot p^2({\infty}).
\ee
where $p_v$ is a constant of motion resulting from the fact the  metric does not depend on the coordinate $v$ (in fact it is the  proportionality coefficient   between  the coordinate $u$ and the affine parameter).
Now,  let us recall that  the BJR coordinates $(u, \hat {\bf x }_{in},\hat v_{in})$   are defined by 
\be
\label{e19x}
{\bf x}=P_{in}\hat {\bf  x}_{in}, \quad
  v=\hat v_{in}-\frac 1 4\hat {\bf x}_{in}\cdot\dot{ \hat{ H}} _{in}\hat {\bf x}_{in},
 \qquad  \hat H_{in}=P_{in}^TP_{in};
\ee
 in terms of them the plane  gravitational  wave takes the form 
 \be
 \label{e19l}
  \hat g_{in}=2dud\hat v_{in}+ d \hat {\bf x}_{in}\cdot \hat H_{in}d \hat {\bf x}_{in}.
 \ee
Let us also note  that  the   solution to the  matrix Sturm-Liouville  differential equation    is strictly related  to  the Killing vectors and  consequently  the isometry groups  of the plane gravitational waves (see e.g.  \cite{b2c,b2d}). 
The  choice   (\ref{e19c}) of the initial conditions  implies that the BJR coordinates  and the metric (\ref{e19l}) coincide   with the Brinkmann ones at minus infinity (and, due to the  asymptotic flatness of  gravitational wave, with  Minkowski one).  On the other hand, the BJR  coordinates  (metrics)  are singular at the point where  the matrix $P_{in}$ vanishes; thus we need  several charts to cover the whole Brinkmann manifold (see e.g.   \cite{b12a,b12b}). Analogously, one can  construct  the  matrix $P_{out}$ around  plus infinity  and consequently  the ``out" BJR coordinates; however,  as we have indicated above   such  two maps may not  be sufficient  to cover the whole   Brinkmann manifold  (see  also \cite{b7}). 
\par
{Now, let us  apply  the above general  considerations to the case of the circularly polarized gravitational waves defined by the profile (\ref{e9}).} First,   we should  rewrite  the initial  conditions in $y$'s variables defined by (\ref{e2}). For the  transverse  velocities, applying L'Hospital's rule,  one obtains the relation 
\be
\label{e19e}
\dot {\bf  x}(\pm\infty)=\pm R(\pm{ \pi}/{ 2}){\bf y}(\pm \pi /2).
\ee
Next, assuming $\dot {\bf x}(-\infty)=0$,  straightforward computations yield 
\be
\label{e19eee}
{\bf x}(-\infty)=\epsilon R(-{ \pi}/{ 2}){\bf y}'(- \pi/ 2).
\ee
Thus the solutions ${\bf y_1}$ and ${\bf y_2}$  to eqs.  (\ref{e17}) with the initial conditions 
\be
\label{e19ee}
{\bf y}_{1,2}(- \pi/ 2)=0, \quad {\bf y}_{1,2}'(- \pi /2)=\frac{1}{\epsilon} R( \pi/ 2){\bf e}_{1,2},
\ee
lead directly to  the  solutions  $ {\bf x_1}$ and ${\bf x_2}$   discussed  above and, consequently, to  the matrix $P_{in}$ (see  eq. (\ref{e19d})).  Moreover, the determinant of the matrix $P_{in}$ can be also expressed in terms of ${\bf y}_{1,2}$, indeed  
\be
\label{e19f}
\det( P_{in}(u))=\frac {\epsilon^2}{\cos^2(\tilde u)}\det\left( R(\tilde u){\bf y}_1(\tilde u),\  R(\tilde u){\bf y}_2(\tilde u) \right)=\frac {\epsilon^2}{\cos^2(\tilde u)}\det\left( {\bf y}_1(\tilde u),\  {\bf y}_2(\tilde u) \right),
\ee
where $\tilde u=\tan^{-1}(u/\epsilon)$. Finally, due to (\ref{e19k})  and (\ref{e19e}) one gets   
\be
\begin{split}
\label{e19g}
\frac 1 J=\det(\dot P_{in}(\infty))= \det (R({ \pi}/{ 2}){\bf y}_1( \pi/ 2),R({ \pi}/{ 2}){\bf y}_2( \pi/ 2))
=\det ({\bf y}_1( \pi/ 2),{\bf y}_2( \pi/ 2)).
\end{split}
\ee
In summary, we express  the basic quantities corresponding to  the circularly polarized plane gravitational pulse $g^{(2)}$    in terms of  solutions to, explicitly integrable,   equations (\ref{e17}). 
\par
{Concluding our considerations, let us  analyse  the Dirac delta limit}. For the  case of $g^{(1)}$  such a  limit (including  the cross section  and  notion of  the conformal group)  was discussed in \cite{b7}.  In the case of $g^{(2)}$   we take $a=\frac{2\epsilon^3}{\pi}$ and  $\gamma=r\epsilon$, i.e. $\omega=r>0$.   Then in the limit $\epsilon\mapsto 0$ one obtains 
\be
\label{e22}
H^{(2)}(u)
\mapsto 
 \left(
\begin{array}{cc}
1&0\\
0&- 1
\end{array}
\right)\delta(u)\cdot
\left\{
\begin{array}{cc}
\frac{\sin(\pi r)}{\pi r(1-r^2)}, &\quad r\neq 1; \\
\frac{1}{2}, &\quad r=1.
\end{array}
\right.
\ee
Thus for the gravitational wave $g^{(2)}$ the Dirac delta limit reduces, up to a constant, to the one  for $g^{(1)}$ (the circular polarization has no time to act),  which coincides with the results obtained in \cite{b16,b14}.
{\subsection{The explicit example}} 
{To illustrate the   results from the above subsection   let us take   the following parameters  $\omega=1$, i.e. $\gamma=\epsilon$ in the circular case.} Then  $ \Omega_+=-\Omega_-=-\Omega$ while eqs.  (\ref{e12}) and  (\ref{e2})  give the following relations
\be
\label{e3}
\begin{split}
 x^1(u)&=\epsilon y^1(\tan^{-1}(  u/ \epsilon)) -uy^2(\tan^{-1}(  u /\epsilon)),\\
 x^2(u)&= uy^1(\tan^{-1}( u/ \epsilon))+\epsilon y^2(\tan^{-1}( u /\epsilon)),
\end{split}
\ee
 where $y$'s contain both the trigonometric and hyperbolic functions. Explicitly 
\be
\label{e4}
\begin{split}
 y^1(\tilde u)&=(\Omega-\lambda^2_+)(C_3\sin(\lambda_+\tilde u)+C_4\cos(\lambda_+\tilde u))+(\lambda_-^2+\Omega)(C_1\sinh(\lambda_-\tilde u)-C_2\cosh(\lambda _-\tilde u)),\\
 y^2(\tilde u)&=2\lambda_+(-C_3\cos(\lambda_+\tilde u)+C_4\sin(\lambda_+\tilde u))+2\lambda_-(C_2\sinh(\lambda_-\tilde u)-C_1\cosh(\lambda _-\tilde u)),
 \end{split}
\ee
where 
\be
\label{e20}
\lambda_\pm=\sqrt{2\left(\pm1+\sqrt{1+\frac{\Omega^2}{4}}\right)}.
\ee
From the above, it is  clear  that freely falling test particles in the neighbourhood of plus null infinity  fly apart along straight lines with the constant velocity i.e. they  exhibit the so-called  velocity memory effect (a test particles initially at rest fly apart with non-vanishing constant velocity after a burst of a gravitational wave has passed, see \cite{b2c,b2d} as well as  \cite{b14} for the circularly polarized case).
\par 
Now, imposing  the initial conditions (\ref{e19ee}) on the solutions  (\ref{e4}) one obtains the following form  of  ${\bf y}_{1,2}$
\be
({\bf y}_1(\tilde u),{\bf y}_2(\tilde u))=\frac{1}{\epsilon(\lambda_-^2+\lambda_+^2)}
\left(
\begin{array}{cc}
-2X(\tilde u)&Y(\tilde u)\\
Z(\tilde u)&-2X(\tilde u)
\end{array}
\right),
\ee
where 
\be
\label{e19m}
\begin{split}
X(\tilde u)&=\cos(\lambda_+(\tilde u+\frac \pi 2))-\cosh(\lambda_-(\tilde u+\frac \pi 2)),\\
Y(\tilde u)&=(\lambda_--\lambda_+)\sin(\lambda_+(\tilde u+\pi/ 2))-(\lambda_+ +\lambda_-)\sinh(\lambda_-(\tilde u+ \pi /2)),\\
Z(\tilde u)&=(\lambda_+ +\lambda_-)\sin(\lambda_+(\tilde u+\pi /2))+(\lambda_--\lambda_+)\sinh(\lambda_-(\tilde u+ \pi/ 2)).
\end{split}
\ee
Thus by virtue  (\ref{e19d})  and (\ref{e3})  one obtains the explicit form of the matrix $P_{in}$  and consequently the transformation rule to the  BJR coordinates.  Moreover one gets  
\be
\label{e19h}
\det ({\bf y}_1(\tilde u),{\bf y}_2(\tilde u))=\frac{\Omega\sinh(\lambda_-(\tilde u+\frac \pi 2))\sin(\lambda_+(\tilde u+\frac \pi 2)) -2\cos(\lambda_+(\tilde u+\frac \pi 2))\cosh(\lambda_-(\tilde u+\frac \pi 2))+2}{\epsilon^2(4+\Omega^2)};
\ee
thus,   due to  (\ref{e19x}), (\ref{e19l})  and (\ref{e19f}),  $\det(P_{in})$ as well as    $\det(\hat g_{in})$ can be directly  computed. 
{In consequence the passage to the BJR coordinates is given by the explicit formulae.}   
\par 
Next,   by virtue of (\ref{e19g}) one immediately gets 
\be
\label{e19j}
J=\frac{\epsilon^2(4+\Omega^2)}{\Omega\sinh(\lambda_- \pi)\sin(\lambda_+ \pi ) -2\cos(\lambda_+ \pi )\cosh(\lambda_- \pi )+2},
\ee
and, consequently,  the classical cross section (\ref{e19i})  (when the final momenta vanish, i.e.  the denominator in    (\ref{e19j}) is zero, the notion of cross section should be  modified). 
\par 
In order to get some insight into the focusing and singularity problems let us  factorize the determinant (\ref{e19h}).  After some computations one gets
\be
\det ({\bf y}_1(\tilde u),{\bf y}_2(\tilde u))=\frac{V(\tilde u)W(\tilde u)}{\epsilon^2(4+\Omega^2)},  
\ee
where 
\be
\begin{split}
V(\tilde u)&=(\lambda_++\lambda_-)\sinh(\frac{ \lambda_-}{2}(\tilde u+\frac \pi 2))\cos(\frac{\lambda_+}{2}(\tilde u+\frac \pi 2))\\
&-(\lambda_--\lambda_+)\cosh(\frac{ \lambda_-}{2}(\tilde u+\frac \pi 2))\sin(\frac{\lambda_+}{2}(\tilde u+\frac \pi 2)),\\
W(\tilde u)&=(\lambda_++\lambda_-)\sin(\frac{ \lambda_+}{2}(\tilde u+\frac \pi 2))\cosh(\frac{\lambda_-}{2}(\tilde u+\frac \pi 2))\\
&+(\lambda_--\lambda_+)\cos(\frac{ \lambda_+}{2}(\tilde u+\frac \pi 2))\sinh(\frac{\lambda_-}{2}(\tilde u+\frac \pi 2)).
\end{split}
\ee
Thus, in agreement with the general theory \cite{b12a},  the BJR coordinates are singular when either  $V$ or $W$ vanishes -- depending on  the wave parameters $a$ and $\epsilon$.  Furthermore,  the focusing holds on  a line (since  $X(\tilde u) \neq 0$, the rank of $P_{in}$ is  at least one).  One can also find the kernel of the matrix $P_{in}$ describing the initial positions for which the  focusing actually appears (by virtue of (\ref{e19r}) the geodesics corresponding to the initial positions differing  by an element of the kernel of  $P_{in}$ will eventually  meet  at  some point). In  our case,  $W(\tilde u_0)=0$, the kernel  of $P_{in}$  is spanned by  the vector
\be
 \Big((\lambda_++\lambda_-)\sin(\frac{ \lambda_+}{2}(\tilde u_0+\frac \pi 2)),\ 2\cos(\frac{ \lambda_+}{2}(\tilde u_0+\frac \pi 2))\Big) R(-\tilde u_0);
\ee
 the case $V(\tilde u_0)=0$ can be dealt with analogously.
\subsection{Some  remarks on the  pp-waves solvability}
Above we  showed  that  for some special plane  gravitational waves  there exist    explicit solutions to the geodesic equations (such a situation   enables one   to  get some insight into  physically  important phenomena); moreover, we noted  that   the notion of the Niederer transformation  provides a very convenient framework for the considered cases.  Although the  explicit solutions are  very exceptional (and call for    some additional assumptions)   the  question arises if we  can  apply this framework  to   obtain different examples of solvable  pp-wave  spacetimes. 
 In what follows we give an  affirmative  answer to this question. More precisely, we show that the Niederer framework can be  used  to produce some pp-wave spacetimes with analytical solutions   of   geodesics equations; at least in some directions  (the full solvability, as we pointed out above  is very exceptional).  To this end let us consider  the pp-wave metric of the form  
\be
g= \mH(u,{\bf x} )du^2+2dudv+d{\bf x}\cdot d{\bf x},
\ee
describing, in general,  a null fluid solution. 
Then the geodesic equations  can be rewritten in the  form 
\begin{align}
\label{x0}
\overset{..}{{\bf x}}&=\frac {{  \nabla} \mH}{2},\\
\label{x00} \overset {..}{v}&=-\frac{\partial_u{\mH} }{2}-\nabla \mH\cdot \dot {{\bf x}}.
\end{align}
Taking into account  $\frac{d x^\mu}{d \tau}\frac{d x_\mu}{d \tau}=-1$  and the relation $u=\frac{p_v}{m}\tau$  one concludes   that   the last equation  can be directly integrated to  the form 
\be
\label {x1}
\dot v=-\frac{{\mH} }{2}-\frac 12{\bf \dot x}^2-\frac{m^2}{2 p_v^2}.
\ee
Moreover,  when the profile $H$ is a homogenous function of degree two in transverse directions i.e.  
\be
\label{x5}
{\bf x}\cdot \nabla  \mH=2\mH,
\ee
 then  eq. \eqref{x1} can be  directly  integrated once more yielding 
\be
v=-\frac{\dot {\bf  x}\cdot  {\bf  x} }{2}-\frac{m^2}{2 p_v^2} u+C,
\ee
where ${\bf x}$ is a solution of  eqs. \eqref{x0}.  In consequence, for  a homogenous function $H$ the solution  problem of the  geodesic equations for pp-waves reduces to the set of eqs. \eqref{x0}; in general case, there remains one integration of the $v$-coordinate.  Thus in what follows we focus on the set of equations given by \eqref{x0}.  
\par
{
Let us take an  {arbitrary} function $\mK(\tilde {\bf x}) $ of two variables and consider the following profile of the pp-wave metric
\be
\label{x2}
\mH(u,{\bf x})= \frac{1}{{u^2+\epsilon^2}}\mK\left(\frac{{\bf x}}{\sqrt{u^2+\epsilon^2}}\right)+\frac{{\bf  x}^2}{2(\epsilon^2+u^2)^2}.
\ee
Then, after some computations, one can show that  eqs. \eqref{x0} in the new coordinates $\tilde {\bf  x}$, defined by  eqs. (\ref{e12}),  take the following form 
\be
\label{x3}
\tilde {\bf x}''=\tilde \nabla \mK.
\ee
The last  set of equations is nothing  more than the two-dimensional Newton equations for which  some explicit solutions are well known; in these cases one  immediately obtains, by means of the Niederer transformation,  the explicit form of solutions to  eqs. \eqref{x0} with the profile \eqref{x2}. 
For example, let us take  
\be
\mK(\tilde{\bf  x})=\frac{a}{(\tilde  x^1)^2}+\frac{b}{(\tilde  x^2)^2};
\ee
then eqs. \eqref{x3} can be explicitly solved and  in consequence   the transversal part  of the geodesic equations \eqref{x0} with the profile 
\be
\mH(u,{\bf x} )=\frac{a}{(  x^1)^2}+\frac{b}{(  x^2)^2}+\frac{{\bf  x}^2}{2(\epsilon^2+u^2)^2},
\ee 
is solvable.
A similar situation holds for other solvable potentials $\mK$'s, as  the P\"oschl-Teller or Morse potentials.   However, as we pointed out   above there remains one integration of  the $v$-coordinate (see eq. \eqref{x1})  to obtain  the full  explicit solvability.  
}
{
\section{Conformal generators and  Ermakov-Lewis  invariants}
To obtain a  wider view of the   integrability  problem of the  discussed   families  of  the plane waves  let us have a  look at them from the  symmetry point of view.   In the standard approach one  considers the isometry group. Then the integrals of motion, associated with the Killing vectors, are well known and can be used to reduce the  solution problem of  the geodesic equations  to some integrals (though, in general, these integrals  cannot be explicitly computed).   The possible isometry  groups of  the vacuum pp-waves were classified, see  \cite{b6a,b6b,b6e} and references therein.  Of course, the most interesting cases should be the  ones with the maximal symmetry; such a  group is at most six-dimensional.  The group of  the dimension six is realized by two, thoroughly   studied,  classes of plane gravitational waves (see  Subsection  \ref{sub1}), whereas the dimension five is  realized by the remaining plane gravitational waves.  As one can expect  for the classes   of  the dimension six (i.e. the maximal dimension of the isometry group)   the geodesic equations can be analytically solved.  
}
\par 
{ 
Now, let us  enlarge the symmetry group  by considering the conformal transformations.  This is interesting since in the case of   pp-waves there can appear additional  integrals of motion. For the geodesic $x^\mu(\tau)$ and  a conformal vector field $K$, $L_K g= 2\psi g$, the following identity (along the geodesic) holds
\be
\label{x4}
\frac{d}{d\tau}(K_{\mu}\frac{dx^\mu}{d\tau})=\psi\frac{dx_\mu}{d\tau}\frac{dx^\mu}{d\tau}=-\psi;
\ee 
on the other hand,  for the  pp-waves the $\tau$  parameter is proportional to $u$. Now,   let us  consider the  homothetic vector field $Y$, then $\psi$ is a constant and thus \eqref{x4} gives a, linearly,   $u$-dependent integral of motion. 
For example, in the case of   plane waves  there always exists  a homothetic vector    of the following form 
\be
\label{x8}
Y=2v\partial_v+{\bf x}\cdot \nabla.
\ee
The above  vector field  is also a  homothetic one for a pp-wave  spacetime   if  and only if the condition \eqref{x5} holds, i.e. when   eq. \eqref{x1}   for $v$ can be directly integrated.  Thus the existence of  the homothetic vector \eqref{x8}  ensures  the  expression of the $v$ coordinate in terms of  solutions to  eqs. \eqref{x0} (if the later ones are analytically solvable then  one gets  the full explicit solvability). 
}
\par 
{
Now, let us focus on the  proper conformal transformations. Then, it turns out that  for the  so-called special  conformal  vectors of the, non-flat, pp-waves the factor $\psi$  is a linear function of $u$ only (see \cite{b6e}). Thus again  one can use eq. \eqref{x4} to find additional integrals of motion containing $u^2$ term. Moreover,  in view of the above  discussion,  it would be desirable  to  have   a rich isometry group  as well as a homothetic vector field. In consequence,  we should  look for  pp-waves  with the maximal conformal group and admitting a proper conformal transformation. For the  vacuum  solutions to the Einstein equations   there  is only one  such class and it is precisely given  by the considered  plane waves, eqs. (\ref{e8}) and (\ref{e9}).  Let us analyse this situation in some detail.
}
\par
{
In the case of the  linearly polarized case,   defined by   the profile $H^{(1)}$ (see  (\ref{e8})), the  conformal vector   field and conformal factor are  of the following form 
\be
\label{x7}
K^{(1)}=(u^2+\epsilon^2)\partial_u-\frac1 2 {\bf x}^2\partial _v+u{\bf x }\cdot \nabla, \quad \psi =u.
\ee
By virtue of \eqref{x4}  the vector field \eqref{x7}   gives  the  integral of motion
\be
I^{(1)}=(u^2+\epsilon^2)({\bf x}\cdot H^{(1)}{\bf x}+\dot v)-\frac1 2 {\bf x}^2+u{\bf x }\cdot \dot {\bf x}+\frac{m^2}{2p_v^2}u^2.
\ee
Using eq. \eqref{x1}  one gets
\be
I^{(1)}=-\frac{m^2\epsilon^2}{2p_v^2}+ a\frac{(x^1)^2-(x^2)^2}{2(u^2+\epsilon^2)}-(u^2+\epsilon^2)\frac{\dot {\bf x}^2}{2}-\frac1 2 {\bf x}^2+u{\bf x }\cdot \dot {\bf x}.
\ee
Now, the key observation is that this integral of motion can be rewritten in terms of the  Ermakov-Lewis  invariants.
To this end let us introduce the function
\be
\label{x6}
\rho(u)=\frac{\sqrt{u^2+\epsilon^2}}{\sqrt{\epsilon}}.
\ee
Then the integral of motion $I^{(1)}$ can be  expressed as follows 
\be
I^{(1)}=-\frac{m^2\epsilon^2}{2p_v^2}-\frac{\epsilon}{2}\left[(\rho \dot x^1-\dot \rho x^1)^2+\frac{\Lambda_1 (x^1)^2}{\rho^2}\right]
-\frac{\epsilon}{2}\left[(\rho \dot x^2-\dot \rho x^2)^2+\frac{\Lambda_2 (x^2)^2}{\rho^2}\right],
\ee
where $\Lambda_i$ are defined by eqs.  (\ref{e11}).  
Moreover, the function $\rho$ satisfies  the set of the  Ermakov-Milne-Pinney  equations  for the profiles $H^{(1)}$  and  $\tilde H^{(1)}$ (the latter one is  defined by the  frequencies $\Lambda_i$, see  eqs. (\ref{e14})  and (\ref{e11})), namely  
  \be
 \overset{..}{\rho}I-H^{(1)}\rho=-\frac{\tilde  H^{(1)}}{\rho^3}.
 \ee
Thus the integral of motion  defined by the  conformal vector field \eqref{x7} is the sum of two Ermakov-Lewis  invariants \cite{a0a}-\cite{a2} with  the frequencies $\Lambda_{1,2}$, respectively. This information can be used to  find the   solutions of  the transverse  part  of the  geodesic equations.  
According to the  general  procedure, see  e.g. \cite{a5},  the transformation
\be
\frac{d \tilde u}{du}=\frac{1}{\rho^2(u)}, \quad {\bf x}=\sqrt{\epsilon}\rho(u)\tilde {\bf x},
\ee
should  relate the $u$-dependent linear oscillator,  defined by $H^{(1)}$, to the harmonic one with  the frequencies  $\Lambda_{1,2}$. In  our case, i.e. $\rho$   given by  \eqref{x6},  the above formulae yield the Niederer transformation (cf. eqs. (\ref{e12}))  and consequently the explicit integrability.   
}
\par
{
Furthermore,  it turns out  (see \cite{a5} and references therein) that    the Ermakov-Lewis  invariants can be interpreted as the ``classical"  energy in the new coordinates $\tilde {\bf x} ,\tilde u$. In our  case this leads to the identity 
\be
I^{(1)}= -\epsilon^2\left[\frac{m^2}{2p_v^2}+ E^{(1)}\right],
\ee
where 
\be
E^{(1)}=\frac12 \tilde {\bf x}'^2-\frac12\tilde {\bf x}\cdot\tilde H^{(1)} \tilde {\bf x}=\frac12 \tilde {\bf x}'^2+ \frac{\Lambda_1(\tilde x^1)^2}{2} + \frac{\Lambda_2(\tilde x^2)^2 }{2}.
\ee
In consequence, we obtain a  more transparent   interpretation of the integral of motion associated with the proper conformal generator $K^{(1)}$.    
}
\par 
{
Now let us  consider the second family of  more interesting plane  gravitational waves which exhibits the maximal conformal symmetry,  i.e.  defined by $H^{(2)}$  (equivalently by $G^{(2)}$ see eqs. (\ref{e9})).  Then, the proper  conformal field $K^{(2)}$  is of the  following form
\be
K^{(2)}=K^{(1)}-{\gamma}(x^2\partial_1-x^1\partial_2),
\ee
with the same conformal factor $\psi=u$.   As in the previous case one can express, by means of \eqref{x6},  the corresponding integral of motion   as follows 
\begin{align}
I^{(2)}&=-\frac{m^2\epsilon^2}{2p_v^2}+ a\frac{{\bf x}\cdot G^{(2)}(u){\bf x}}{2(u^2+\epsilon^2)}-(u^2+\epsilon^2)\frac{\dot {\bf x}^2}{2}-\frac1 2 {\bf x}^2+u\dot {\bf x }\cdot  {\bf x}-{\gamma} \dot {\bf x}\times {\bf x}=\\
&-\frac{m^2\epsilon^2}{2p_v^2}-\frac \epsilon 2 \left[(\rho \dot{\bf x}-{\bf x} \dot \rho)^2-\frac{{\bf x}}{\rho}\cdot( \frac{a}{\epsilon^2}G^{(2)}(u)-I)\frac{\bf x}{\rho} \right]-{\gamma} \dot {\bf x}\times {\bf x}.
\end{align}
In contrast to the  previous case this time the terms in the square brackets do not form the  Ermakov-Lewis invariants (since  the matrix $G^{(2)}$ depends explicitly on $u$, only together with the last term $I^{(2)}$ is an integral of motion).  However, the function $\rho$  satisfies    Ermakov-Milne-Pinney  type equation
 \be
 \overset{..}{\rho}I-H^{(2)}\rho=-\frac{\tilde  H^{(2)}}{\rho^3};
 \ee
thus, in agreement with our previous considerations, it leads to the Niederer transformation and consequently to  eqs. (\ref{e14b}); the solvability is obtained  by further  transformation to the $y$'s coordinates, see eqs. (\ref{e2}).
In view of the above,  we expect that the  integral of motion $I^{(2)}$  should be related  to   total energy for the system  defined by the equations  of motion (\ref{e17}). Indeed, one can check that the integral of motion associated with the conformal vector $K^{(2)}$ is  of the form
\be
I^{(2)}=-\epsilon^2\left[\frac{m^2}{2p_v^2}+ E^{(2)}\right],
\ee
where 
\be
E^{(2)}= \frac12 ({\bf y}')^2+\frac 12\Omega_+(y^1)^2+\frac 12 \Omega_- (y^2)^2,
\ee
and  $\Omega_\pm$ are given by eqs. (\ref{e18}). 
In summary, we showed that  both homothetic and conformal transformations can be crucial in the solvability of geodesic equations; moreover, the integrals associated with  proper conformal generators can be expressed as the  total energy  of the   harmonic oscillators described   by  the frequencies  directly related to parameters defining  the gravitational waves. 
}
\section{Conclusions and  outlook}
Let us summarize.  In the present work  we showed that the Niederer transformation can  explicitly connect   time-dependent linear oscillators  with   the ordinary harmonic  ones (including isotropic and anisotropic cases).   A geometric interpretation of this situation is provided by   the special families (strictly related to the proper  conformal transformations)   of  conformally flat generalized plane  wave
spacetimes  in the isotropic case, and by   exact gravitational   waves in the anisotropic case. Such an observation allows us  to show  directly that the     plane gravitational  waves, corresponding to the  geodesically complete manifolds and exhibiting   the  maximal, 7-dimensional,  conformal symmetry,  admit  analytical solutions  of the geodesics equations.  This   enables one to understand better the  interaction (scattering)  of  particle  with the  gravitational fields (the classical cross section, singularity and  velocity memory effect). Moreover,  in the  agreement with  results obtained  in Ref. \cite{b14} we showed  that the  Dirac delta  limit of  the circularly polarized case reduces to that for the  linearly polarized  one.
{Finally,  the role  and meaning of   the  additional integrals of motions  associated with  the conformal generators were discussed  by means of the  Ermakov-Lewis  invariants and   their  more transparent interpretation was obtained.
}
 \par
The results   obtained can be extended in various directions. First,  it would be interesting  to consider their quantum counterparts, including both the  time-dependent linear oscillator  as well as the gravitational case  (e.g. \cite{a5, b12b,b19}).  Next, the analytical solutions can give   a better insight into singularity problems     of the BJR coordinates, e.g. \cite{x0b,b7}.  Such solutions can be also  useful in the recent studies concerning some aspects of the energy transfer \cite{b21a,b21b} as well as in the context of some optical effects in nonlinear plane  gravitational waves \cite{b22a,b22b}.  
{Furthermore, let us recall that the Penrose limit of spacetimes yields the plane
gravitational waves; thus, the question is which spacetimes  correspond to the conformally  distinguished ones  (cf.  \cite{x0c}).  This is of some importance  due to the recent result   that the  memory is encoded in the   Penrose  plane wave limit of the original gravitational wave.} 
\par 
Moreover, one of the most interesting topics of  further investigations is a construction of electromagnetic backgrounds important for the light-matter interaction.  This is motivated by the recent results on the classical double  copy approach \cite{b8a}-\cite{b8d}  which enable one  to  construct  some  electromagnetic fields from  solutions to the Einstein equations.  One of the manifestation of  the  double copy   is the observation  that the plane gravitational waves not only satisfy  the vacuum  Einstein equations but also the wave equation; thus they can be used to construct some potentials of  the  electromagnetic fields.  According to this approach,   with  the metric  $g$, see  eq. (\ref{e00e}), one  relates  the following electromagnetic one-form  (in the light-cone coordinates)
\be
\label{e23}
\mA= -{\bf x}\cdot H(u){\bf x}du .
\ee
When   $\textrm{tr}(H)=0$, i.e. in the case  of  the   plane  gravitational wave, the above potential give the electromagnetic field    $\vec E $ and $\vec B$   which  satisfies the  vacuum  Maxwell equations. However, in contrast to its gravitational counterpart,  such a field is not, in general, a plane   electromagnetic wave.  
Furthermore,     electromagnetic field obtained in this way   is the pure radiation (the energy density and the Poynting vector form a null four-vector or   equivalently the square of  Riemann-Silberstein vector vanishes).  Thus  it  can describe a vortex of electromagnetic field \cite{b16a}.  An  important example of such  a situation is a vortex proposed in Ref. \cite{b9} which  can act as a beam guide for charged particles; moreover, it  provides analytically solvable example   and  is an approximation to more realistic beams \cite{b9b}.  Recently, it has been  discussed  in Ref. \cite{b18} in the context of the above  mentioned  gauge--gravity duality.  
In view of our  results  one can expect that the electromagnetic backgrounds corresponding to polarized gravitational waves defined by the profiles (\ref{e8})  and (\ref{e9})  also give analytically solvable electromagnetic backgrounds with vortices (cf.  results of  Ref. \cite{b23d,b23e}); this is evident  as far as  the transverse directions are considered because then the Lorentz equations (in the light-cone coordinates)  coincide, up to a constant, with the geodesic equations, see e.g. \cite{b18}. Moreover, the recently discussed models  which are important for the light-matter interaction, in particular for  the strong focusing of  intense laser pulses \cite{b10a,b10b},  should   be also obtainable in a similar way; subsequently, they could be extended by adding  the new solvable examples or focusing criteria. 
{Some of these problems are under the consideration  and    preliminary results are presented  in \cite{x11}.}
\vspace{0.5cm}
\par
{\bf Acknowledgments}
\par
The  authors would like to thank    Piotr  Kosi\'nski for helpful  discussions as well as    Peter  Horvathy  for valuable  suggestions and remarks.   Comments of  Cezary Gonera,  Joanna Gonera and Pawe\l   \ Ma\'slanka  are  also  acknowledged. 
{We are also  grateful to the anonymous referees for their stimulating  questions and suggestions.} This   work  has   been partially   supported   by   the   grant  2016/23/B/ST2/00727  of  National  Science  Centre, Poland. 
 
\end{document}